\def\mm{{\mu\text{m}}}

\documentclass[aps,prappl,twocolumn,groupedaddress,showkeys,showpacs,superscriptaddress,floatfix]{revtex4-1}
\usepackage{epsfig}
\usepackage{multirow}
\usepackage{amsmath, amssymb,mathtools}
\usepackage{color}
\usepackage{graphicx}
\usepackage{dcolumn}
\usepackage{bm}
\usepackage{subfigure}
\usepackage[utf8]{inputenc}
\usepackage[T1]{fontenc}
\usepackage{tikz}

\begin{document}

\title{Solitonic Josephson thermal transport}

\author{Claudio Guarcello}
\email{claudio.guarcello@nano.cnr.it}
\affiliation{NEST, Istituto Nanoscienze-CNR and Scuola Normale Superiore, Piazza San Silvestro 12, I-56127 Pisa, Italy}
\affiliation{Radiophysics Department, Lobachevsky State University, Gagarin Avenue 23, 603950 Nizhni Novgorod, Russia}
\author{Paolo Solinas}
\affiliation{SPIN-CNR, Via Dodecaneso 33, 16146 Genova, Italy}
\author{Alessandro Braggio}
\affiliation{NEST, Istituto Nanoscienze-CNR and Scuola Normale Superiore, Piazza San Silvestro 12, I-56127 Pisa, Italy}
\author{Francesco Giazotto}
\affiliation{NEST, Istituto Nanoscienze-CNR and Scuola Normale Superiore, Piazza San Silvestro 12, I-56127 Pisa, Italy}

\date{\today}

\begin{abstract}
We explore the coherent thermal transport sustained by solitons through a long Josephson junction, as a thermal gradient across the system is established. We observe that a soliton causes the heat current through the system to increase. Correspondingly, the junction warms up in correspondence of the soliton, with temperature peaks up to, e.g., approximately $56\;\text{mK}$ for a realistic Nb-based proposed setup at a bath temperature $T_{bath}=4.2\;\text{K}$. The thermal effects on the dynamics of the soliton are also discussed. Markedly, this system inherits the topological robustness of the solitons. In view of these results, the proposed device can effectively find an application as a superconducting thermal router in which the thermal transport can be locally mastered through solitonic excitations, which positions can be externally controlled through a magnetic field and a bias current.
\end{abstract}


\maketitle

\section{Introduction}
\label{Intro}\vskip-0.2cm

The physics of coherent excitations has relevant implications in the field of condensed matter. Such coherent objects emerge in several extended systems and are usually characterized by remarkable particle-like features. In the past decades, these notions played a crucial role for understanding various issues in different areas of the physics of continuous and discrete systems~\cite{Sco99,Dau06}. 
A Josephson junction (JJ) is a model system to appreciate coherent excitations, and, specifically, a superconductor-insulator-superconductor (SIS) long JJ (LJJ) is the prototypal solid-state environment to explore the dynamics of a peculiar kind of solitary waves, called soliton~\cite{Par93,Ust98}. These excitations give rise to readily measurable physical phenomena, such as step structures in the I-V characteristic of LJJs and microwaves radiation emission. Moreover, a soliton has a clear physical meaning in the LJJ framework, since it carries a quantum of magnetic flux, induced by a supercurrent loop surrounding it, with the local magnetic field perpendicularly oriented with respect to the junction length~\cite{McL78}. Thus, solitons in the context of LJJs are usually referred to as fluxons or Josephson vortices. Measured for the first time more than 40 years ago~\cite{Sco69,Ful73}, LJJs are still nowadays an active research field~\cite{Ooi07,Gul07,And10,Lik12,Mon12,Gua13,Mon13,Gra14,Cue14,Kos14,Val14,Sol15,Gua15,Vet15,Pan15,GuaValSpa16,Gol17,GuaSol17,Hil18}. Indeed, the fact that a single topologically protected excitation, i.e., a flux quantum, can be moved and controlled by bias currents, created by the magnetic field, manipulated through shape engineering~\cite{Gol01,Car02,CarMar02,Gul07,Dob09}, or pinned by inhomogeneities~\cite{Ust91,Feh92}, naturally stimulated a profusion of ideas and applications. 

Practically, several electric and magnetic features concerning solitons in LJJs were comprehensively hitherto explored, but little is known about the soliton-sustained coherent thermal transport through a temperature-biased junction. This issue falls into the emerging field of coherent caloritronics~\cite{Gia06,MarSol14,ForGia17}, which deals with the manipulation of heat currents in mesoscopic superconducting devices. Here, the aim is to design and realize thermal components able to master the energy transfer with a high degree of accuracy. In this regard, we propose to lay the foundation of a new branch of fast coherent caloritronics based on solitons, with the end to build up new devices exploiting this highly-controllable, “phase-coherent” thermal flux. Specifically, the feasibility of using a LJJ as a thermal router~\cite{Tim17}, in which thermal transport can be locally handled through solitonic excitations, is very promising.

\begin{figure*}[!!ht]
\includegraphics[width=\textwidth]{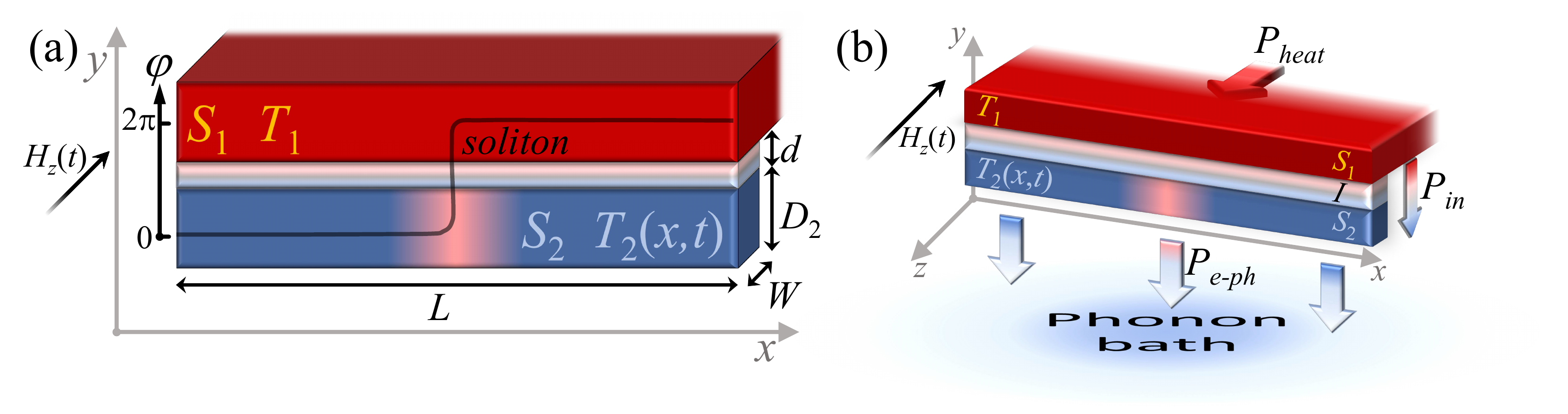}
\caption{\textbf{a}, A superconductor-insulator-superconductor (SIS) rectangular long Josephson junction (LJJ) excited by an external in-plane magnetic field $H_{z}(t)$. The length and the width of the junction are $L\gg\lambda_{_{J}}$ and $W \ll\lambda_{_{J}}$, respectively, where $\lambda_{_{J}}$ is the Josephson penetration depth. Moreover, the thickness $D_2 \ll\lambda_{_{J}}$ of the electrode $S_2$ is indicated. A soliton within the junction, corresponding to a 2$\pi$-twist of the phase $\varphi$, is represented. $T_i$ is the temperature of the superconductor $S_i$ and $d$ is the insulating layer thickness. \textbf{b}, Thermal model of the device, as the thermal contact with a phonon bath is taken into account. The heat current, $\mathcal{P}_{in}$, flowing through the junctions depends on the temperatures and the solitons eventually set along the system. $P_{e-ph}$ represents the coupling between quasiparticles in $S_2$ and the lattice phonons residing at $T_{bath}$, whereas $P_{heat}$ denotes the power injected into $S_1$ through heating probes in order to impose a fixed quasiparticle temperature $T_1$. The arrows indicate the direction of heat currents for $T_1>T_2>T_{bath}$. }
\label{Fig01}
\end{figure*}

After the earlier prediction in 1965 by Maki and Griffin~\cite{Mak65}, only recently phase-coherent thermal transport in temperature-biased Josephson devices has been confirmed experimentally in several interferometer-like structures~\cite{GiaMar12,Gia12,Gia13,Mar14,ForBla16,For17}. The thermal modulation induced by the external magnetic field was demonstrated in superconducting quantum-interference devices (SQUID)~\cite{GiaMar12,Gia12} and short JJs~\cite{Gia13,Mar14}. Furthermore, in LJJs the heat current diffraction patterns in the presence of an in-plane external magnetic field have been discussed theoretically~\cite{Gua16}. However, until now no efforts have been addressed to explore how thermal transport across a LJJ is influenced by solitons eventually set along it. Nonetheless, it has been demonstrated theoretically that the presence of a fluxon threading a temperature-biased inductive SQUID modifies thermal transport and affects the steady temperatures of a floating electrode of the device~\cite{Gua17,GuaSolBra18}. Similarly, we demonstrate theoretically that a fluxon arranged within a LJJ locally affects, in a fast timescale, the thermal evolution of the system, and, at the same time, we discuss how the temperature gradient affects the soliton dynamics. 
Finally, being solitons, namely, remarkably stable and robust objects~\cite{Bis80}, at the core of its operation, this system provides an intrinsic topological protection on thermal transport. 

The paper is organized as follows. In Sec.~\ref{PhaseModel}, the theoretical background used to describe the phase evolution of a magnetically-driven LJJ is discussed. In Sec.~\ref{ThermalModel}, the thermal balance equation and the heat currents are introduced. In Sec.~\ref{Results}, the evolution of the temperature of the floating electrode is studied, as a thermal gradient across the system is taken into account. In Sec.~\ref{Conclusions}, conclusions are drawn.

\section{Phase dynamics}
\label{PhaseModel}\vskip-0.2cm

In Fig.~\ref{Fig01}a long and narrow SIS Josephson junction, in the so-called overlap geometry, formed by two superconducting electrodes $S_1$ and $S_2$ separated by a thin layer of insulating material with thickness $d$ is represented. We consider an extended junction with both the length and the width larger than $d$ (namely, $W,L\gg d$). In the geometry depicted in Fig.~\ref{Fig01}, the junction area $A=W\,L$ extends in the $xz$-plane, the electric bias current is eventually flowing in the $y$ direction, and the external magnetic field is applied in the $z$ direction. The thickness of each superconducting electrode is assumed larger than the London penetration depth $\lambda_{L,i}$ of the electrodes material. Since the applied field penetrates the superconducting electrodes up to a thickness given by the London penetration depth, an effective magnetic thickness of the junction $t_d=\lambda_{L,1}+\lambda_{L,2}+d$ can be defined. If $\lambda_{L,i}$ are larger than the thickness of the electrodes $D_i$, the effective magnetic thickness has to be replaced by $\tilde{t}_d=\lambda_{L,1}\tanh\left ( D_1/2\lambda_{L,1} \right )+\lambda_{L,2}\tanh\left ( D_2/2\lambda_{L,2} \right )+d$~\cite{Gia13,Mar14}. 
In the presence of an external in-plane magnetic field $\mathbf{H} (\mathbf{r},t)=(0,0,-H(t)\;\widehat{\mathbf{z}})$, the phase $\varphi$, namely, the phase difference between the wavefunctions describing the carriers in the superconducting electrodes, changes according to $\partial \varphi(x,t)/\partial x=\frac{2\pi}{\Phi_0}\mu_0t_dH(t)$~\cite{Bar82},
where $\Phi_0= h/2e\simeq2\times10^{-15} \textup{Wb}$ is the magnetic flux quantum (with $e$ and $h$ being the electron charge and the Planck constant, respectively), and $\mu_0$ is the vacuum permeability. 
For a long and narrow junction, we assume that $W\ll \lambda_{_{J}}$ and $L\gg \lambda_{_{J}}$, where we introduced the length scale $\lambda_{_{J}}=\sqrt{\frac{\Phi_0}{2\pi \mu_0}\frac{1}{t_d J_c}}$ called \emph{Josephson penetration depth}, where $J_c=I_c/A$ is the critical current area density. Then, in normalized units, the linear dimensions of the junction read $\mathcal{L}=L/\lambda_{_{J}}\gg1$ and $\mathcal{W}=W/\lambda_{_{J}}\ll 1$.

The electrodynamics of a LJJ is usually described by a partial differential equation for the order parameter phase difference $\varphi$, namely, the perturbed sine-Gordon (SG) equation, that in the normalized units $\widetilde{x}=x/\lambda_J$ and $\widetilde{t}=\omega_pt$, with $\omega_p=\sqrt{\frac{2\pi}{\Phi_0}\frac{I_c}{C}}$ being the \emph{Josephson plasma frequency}~\cite{Bar82}, reads~\cite{Lom82,Bar82} 
\begin{equation}
\frac{\partial^2 \varphi(\widetilde{x},\widetilde{t}) }{\partial {\widetilde{x}}^2} -\frac{\partial^2 \varphi(\widetilde{x},\widetilde{t}) }{\partial {\widetilde{t}}^{2}}- \sin\big( \varphi\left ( \widetilde{x},\widetilde{t} \right ) \big )= \alpha\frac{\partial \varphi(\widetilde{x},\widetilde{t}) }{\partial \widetilde{t}}.
\label{SGeq}
\end{equation}
The boundary conditions of this equation takes into account the normalized external magnetic field $\mathcal{H}(t)=\frac{2\pi}{\Phi_0\mu_0}t_d\lambda_{_{J}} H(t)$
\begin{equation}
\frac{d\varphi(0,t) }{d\widetilde{x}} = \frac{d\varphi(L,t) }{d\widetilde{x}}= \mathcal{H}(t).
\label{bcSGeq}
\end{equation}
In Eq.~\eqref{SGeq}, $\alpha=(\omega_p RC)^{-1}$ is the damping parameter (with $R$ and $C$ being the total normal resistance and capacitance of the JJ).

The SG equation admits topologically stable travelling-wave solutions, called \emph{solitons}~\cite{Par93,Ust98}, corresponding to 2$\pi$-twists of the phase (see Fig.~\ref{Fig02}).
For the unperturbed SG equation, i.e., $\alpha=0$ in Eq.~\eqref{SGeq}, solitons have the simple analytical expression~\cite{Bar82}
\begin{equation}
\varphi(\widetilde{x}-u\widetilde{t})=4\arctan \left \{ \exp \left [ \pm \frac{\Big(\widetilde{x}-\widetilde{x}_0-u\widetilde{t} \Big )}{\sqrt{1-u^2}} \right ] \right \},
\label{SGkink}
\end{equation}
where the sign $\pm$ is the polarity of the soliton and $u$ is the soliton speed normalized to the Swihart’s velocity~\cite{Bar82}, namely, the largest group propagation velocity of the linear electromagnetic waves in long junctions.
The moving soliton corresponds to a time variations of the phase, which generates a local voltage drop according to $V(x,t)=\Phi_0/(2\pi)\dot{\varphi}(x,t)$. 

For the numerical simulation of the soliton dynamics, we modelled the normalized external magnetic field $\mathcal{H}(t)$ as a Gaussian pulse exciting the junction end in $x=0$. Accordingly, the boundary conditions become
\begin{equation}
\frac{d\varphi(0,t) }{d\widetilde{x}} = \mathcal{H}(t)\qquad\text{and}\qquad\frac{d\varphi(L,t) }{d\widetilde{x}}=0.
\label{bcSGeq1}
\end{equation}

For simplicity, in our model, i.e., Eq.~\eqref{SGeq}, both the terms $\beta \frac{\partial \varphi}{\partial \widetilde{x}^{2}\partial \widetilde{t}}$~\cite{Lom82,Par93} (with $\beta =\omega_p L_P/R_P$, where $L_P=\mu_0 t_d/W$ and $R_P$ represents scattering of quasiparticles in the superconducting surface layers) and $\Delta_c \frac{\partial H}{\partial \widetilde{x}}$~\cite{Gro92,Par93} (with $\Delta_c$ being a coupling constant) are not included. These terms account for the dissipation due to the surface resistance of the superconducting electrodes and for the spatial gradient of the magnetic field along the junction, respectively. We neglect these contributes since we are interested only to look the interplay between a soliton and the thermal effects resulting from its presence along the system as a temperature gradient across the junction is imposed. In this regard, also the specific mechanism used to excite a soliton is not so relevant. In fact, in the place of a moving soliton generated by a magnetic pulse, we can alternatively design the local control of thermal flux through configurations of steady solitons excited in specific points of the junction via a slowly-varying external magnetic drive applied to both edges of the device~\cite{Gua16}. In this manner, the positions of the solitons are directly dependent on the boundary conditions. Anyway, we observe that, still in this case, a dynamical treatment is crucial for the realistic description of the manipulation of the system, and it leads to peculiar results, such us the hysteresis and the trapping of fluxons~\cite{Gua16}. Alternatively, in an annular geometry~\cite{Dav86}, i.e., a ``closed'' LJJ folded back into itself in which solitons move undisturbed, i.e., without interaction with borders, fluxons can be excited at will~\cite{Ust92,Ust02}, allowing highly-controlled soliton dynamics.

Below, we will briefly discuss also the possibility to control the soliton position by an applied bias current. This feasibility adds an external control knob, making this device more interesting for practical applications.

\section{Thermal effects}
\label{ThermalModel}\vskip-0.2cm

The aim of this section is to explore the thermal flux through the junction, as a soliton is set and a temperature gradient across the junction is imposed. 
Specifically, we observe the evolution of the temperature $T_2(x,t)$, which depends on all the energy local relaxation mechanisms occurring in the electrode $S_2$ (see Fig.~\ref{Fig01}b). 
For the sake of simplicity, we assume that the electrode $S_1$ resides at a fixed temperature $T_1$, which is maintained by the good thermal contact with heating probes. The electrode $S_2$ is in thermal contact also with a phonon bath at temperature $T_{bath}\leq T_2< T_{1}$. 

A characteristic length scale for the thermalization in the diffusive regime can be estimated as the inelastic scattering length $\ell_{in}=\sqrt{D\tau_s}$, where $D=\sigma_N/(e^2N_F)$ is the diffusion constant (with $\sigma_N$ and $N_F$ being the electrical conductivity in the normal state and the density of states at the Fermi energy, respectively) and $\tau_s$ is the recombination quasiparticle lifetime~\cite{Kap76}. For Nb at $4.2\;\text{K}$, one obtains
 $\ell_{in}\sim0.3\;\mm$, namely, a value well below the dimension of a soliton, $\ell_{in}\ll\lambda_{_{J}}$, since $\lambda_{_{J}}\gtrsim6\;\mu\text{m}$ for the device considered here below. When only the length of $S_2$ is much larger than $\ell_{in}$, i.e., $L\gg \ell_{in}$ (namely, the so-called \emph{quasiequilibrium limit}~\cite{Gia06}), the electrode $S_2$ can be modelled as a one-dimensional diffusive superconductor at a temperature varying along $L$. 
 
For the sake of readability, hereafter we will adopt in equations the abbreviated notation in which the $x$ and $t$ dependences are left implicit, namely, $T_2=T_2(x,t)$, $\varphi=\varphi(x,t)$, and $V=V(x,t)$. Then, the evolution of the temperature $T_2$ is given by the time-dependent diffusion equation 
\begin{equation}
\frac{\mathrm{d} }{\mathrm{d} x}\left [\kappa( T_2 ) \frac{\mathrm{d} T_2}{\mathrm{d} x} \right ]+\mathcal{P}_{tot}\left ( T_1,T_2,\varphi \right )=c_v(T_2)\frac{\mathrm{d} T_2}{\mathrm{d} t},
\label{ThermalBalanceEq}
\end{equation}
where the rhs represents the variations of the internal energy density of the system, and the lhs terms indicate the spatial heat diffusion, taking into account the inhomogeneous electronic heat conductivity, $\kappa(T_2)$, and the total heat flux density in the system, namely,
\begin{equation}
\mathcal{P}_{tot}\left ( T_1,T_2,\varphi \right )=\mathcal{P}_{in}\left ( T_1,T_2,\varphi,V \right )-\mathcal{P}_{e-ph,2}\left ( T_2,T_{bath}\right ).
\label{TotalPower}
\end{equation}
This term consists of the incoming, i.e., $\mathcal{P}_{in}\left ( T_1,T_2,\varphi,V \right )$, and outgoing, i.e., $\mathcal{P}_{e-ph,2}\left ( T_2,T_{bath}\right )$, thermal power densities in $S_2$. We stress that the phase dynamics is essential, through $\mathcal{P}_{in}$, to determine the heat flows and the temperature evolution. Therefore, both Eqs.~\eqref{SGeq} and~\eqref{ThermalBalanceEq} have to be solved numerically self-consistently to thoroughly explore the thermal behaviour of the system.

In Eq.~\eqref{TotalPower}, the heat current density $\mathcal{P}_{in}( T_1,T_2,\varphi,V)$ flowing from $S_1$ to $S_2$ is
\begin{eqnarray}\label{Pt}\nonumber
\mathcal{P}_{in}( T_1,T_2,\varphi,V)=&&\mathcal{P}_{qp}( T_1,T_2,V)-\cos\varphi \;\mathcal{P}_{\cos}( T_1,T_2,V)\\
&&+\sin\varphi\; \mathcal{P}_{\sin}( T_1,T_2,V),
\end{eqnarray}
and contains the interplay between Cooper pairs and quasiparticles in tunneling through a JJ predicted by Maki and Griffin~\cite{Mak65}. In fact, $\mathcal{P}_{qp}$ is the heat flux density carried by quasiparticles and represents an incoherent flow of energy through the junction from the hot to the cold electrode~\cite{Mak65,Gia06,Fra97}. Instead, the ``anomalous'' terms $\mathcal{P}_{\sin}$ and $\mathcal{P}_{\cos}$ determine the phase-dependent part of the heat current originating from the energy-carrying tunneling processes involving, respectively, Cooper pairs and recombination/destruction of Cooper pairs on both sides of the junction.
In the adiabatic regime~\cite{Gol13}, the quasi-particle and the anomalous heat current densities, $\mathcal{P}_{qp}$, $\mathcal{P}_{\cos}$, and $\mathcal{P}_{\sin}$ read, respectively,~\cite{Mak65,Gol13,VirSol17}
\begin{widetext}
\begin{eqnarray}\label{Pqp}
\mathcal{P}_{qp}(T_1,T_2,V)&=&\frac{1}{e^2R_aD_2}\int_{-\infty}^{\infty} d\varepsilon \mathcal{N}_1 ( \varepsilon-eV ,T_1 )\mathcal{N}_2 ( \varepsilon ,T_2 )(\varepsilon-eV) [ f ( \varepsilon-eV ,T_1 ) -f ( \varepsilon ,T_2 ) ],\\\label{Pcos}
\mathcal{P}_{\cos}( T_1,T_2,V )&=&\frac{1}{e^2R_aD_2}\int_{-\infty}^{\infty} d\varepsilon \mathcal{N}_1 ( \varepsilon-eV ,T_1 )\mathcal{N}_2 ( \varepsilon ,T_2 )\frac{\Delta_1(T_1)\Delta_2(T_2)}{\varepsilon}[ f ( \varepsilon-eV ,T_1 ) -f ( \varepsilon ,T_2 ) ],\\
\mathcal{P}_{\sin}(T_1,T_2,V)&=&\frac{eV}{2\pi e^2R_aD_2}\iint_{-\infty}^{\infty} d\epsilon_1d\epsilon_2 \frac{\Delta_1(T_1)\Delta_2(T_2)}{E_2}\left [\frac{1-f(E_1,T_1)-f(E_2,T_2)}{\left ( E_1+E_2 \right )^2-e^2V^2}+\frac{f(E_1,T_1)-f(E_2,T_2)}{\left ( E_1-E_2 \right )^2-e^2V^2}\right ],
\end{eqnarray}
\end{widetext}
where $R_a=RA$ is the resistance per area of the junction, $E_j=\sqrt{\epsilon_j^2+\Delta_j(T_j)^2}$, $f ( E ,T )=1/\left (1+e^{E/k_BT} \right )$ is the Fermi distribution function, and $\mathcal{N}_j\left ( \varepsilon ,T \right )=\left | \text{Re}\left [ \frac{ \varepsilon +i\gamma_j}{\sqrt{(\varepsilon +i\gamma_j) ^2-\Delta _j\left ( T \right )^2}} \right ] \right |$ is the reduced superconducting density of state, with $\Delta_j\left ( T_j \right )$ and $\gamma_j$ being the BCS energy gap and the Dynes broadening parameter~\cite{Dyn78} of the $j$-th electrode, respectively.

\begin{figure}[!!b]
\includegraphics[width=\columnwidth]{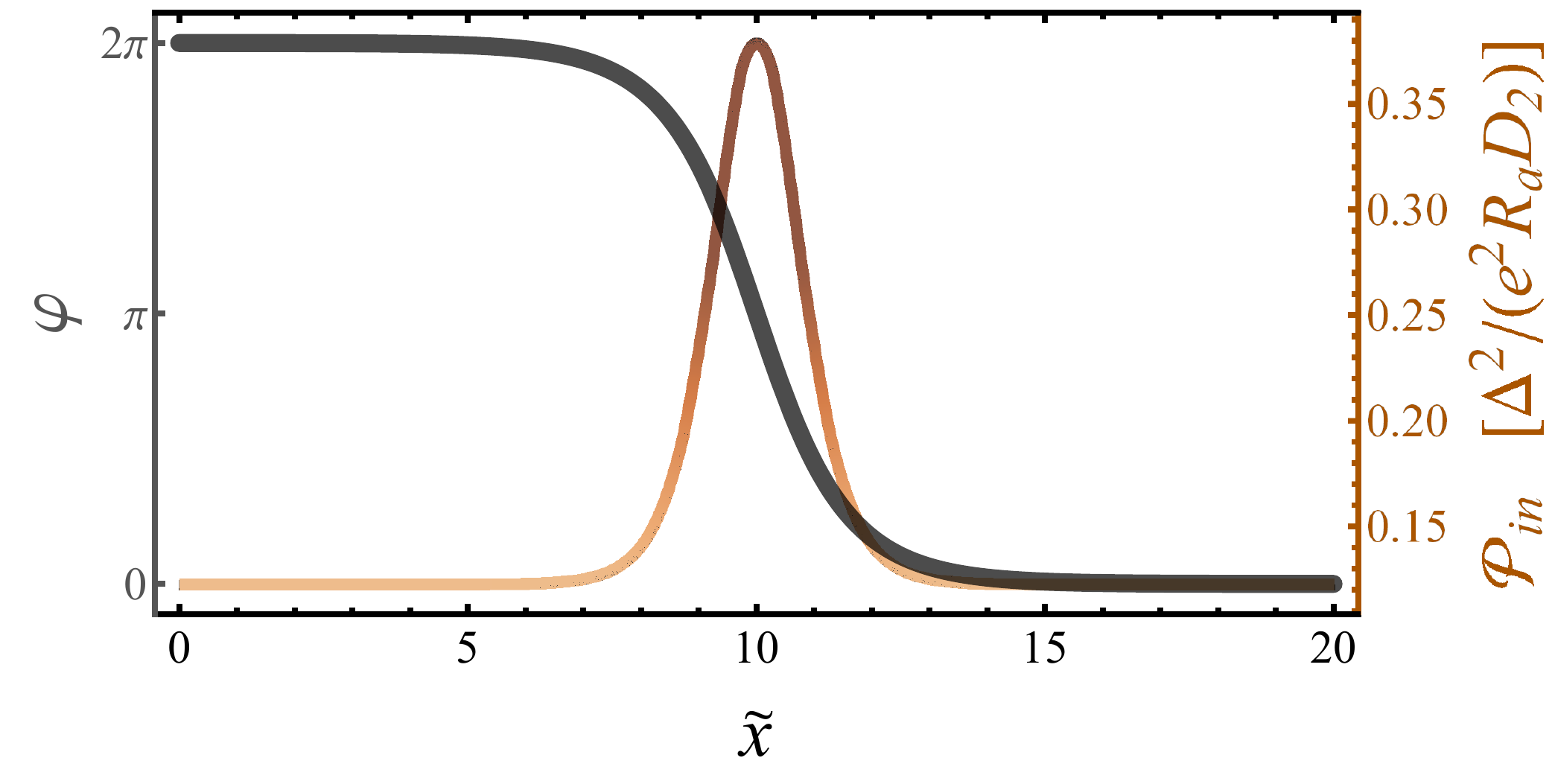}
\caption{Phase profile $\varphi$ (left vertical scale, black line) and the heat power density $\mathcal{P}_{in}(T_1,T_2,\varphi)$ (in units of $\Delta_2^2(0)/(e^2R_aD_2)$), see Eq.~\eqref{Pt} (right vertical scale, orange line), for $T_1=7\;\text{K}$ and $T_2=4.2\;\text{K}$, as a function of the normalized position $\widetilde{x}$, when a steady unperturbed soliton, see Eq.~\eqref{SGkink} for $u=0$, is located in the midpoint of a junction with normalized length $\mathcal{L}=20$.}
\label{Fig02}
\end{figure}

Interestingly, if we calculate the values of the heat current density $\mathcal{P}_{in}( T_1,T_2,\varphi)$ in the presence of a steady unperturbed soliton, described by Eq.~\eqref{SGkink} for $u=0$, an enhancement of $\mathcal{P}_{in}$ just in correspondence of the soliton is observed (see Fig.~\ref{Fig02} assuming for simplicity an homogeneous temperature profile with $T_1=7\;\text{K}$ and $T_{bath}=4.2\;\text{K}$).  Correspondingly, in the presence of a thermal gradient, we expect in the stationary regime a soliton to induce a local warming-up in $S_2$. The peaked shape of $\mathcal{P}_{in}$ shown in Fig.~\ref{Fig02} results from the $\varphi$-dependence of the anomalous contribute $\mathcal{P}_{\cos}$ in Eq.~\eqref{Pt} (notably, the anomalous term $\mathcal{P}_{\sin}$ vanishes in the stationary case, i.e., $\dot{\varphi}=0$). In fact, the coefficient $-\cos\varphi$, that multiplies the $\mathcal{P}_{\cos}$ term, tends to $-1$ for $\varphi\to\{0,2\pi\}$, and it is $+1$ for $\varphi=\pi$, namely, in correspondence of the center of the soliton. Nevertheless, the quasiparticle contribute $\mathcal{P}_{qp}$ represents a positive offset that makes $\mathcal{P}_{in}$ still positive, so that the total heat current flows however from the hot to the cold reservoir. 

In Eq.~\eqref{TotalPower}, the energy exchange between electrons and phonons in the superconductor is accounted by $\mathcal{P}_{e-ph,2}$, which reads~\cite{Pek09}
\begin{eqnarray}\label{Qe-ph}\nonumber
\mathcal{P}_{e-ph,2}&=&\frac{-\Sigma}{96\zeta(5)k_B^5}\int_{-\infty }^{\infty}dEE\int_{-\infty }^{\infty}d\varepsilon \varepsilon^2\textup{sign}(\varepsilon)M_{_{E,E+\varepsilon}}\\\nonumber
&\times& \Bigg\{ \coth\left ( \frac{\varepsilon }{2k_BT_{bath}}\right ) \left [ \mathcal{F}(E,T_2)-\mathcal{F}(E+\varepsilon,T_2) \right ]\\
&-&\mathcal{F}(E,T_2)\mathcal{F}(E+\varepsilon,T_2)+1 \Bigg\},
\end{eqnarray}
where $\mathcal{F}\left ( \varepsilon ,T_2 \right )=\tanh\left ( \varepsilon/2 k_B T_2 \right )$, $M_{E,{E}'}=\mathcal{N}_i(E,T_2)\mathcal{N}_i({E}',T_2)\left [ 1-\Delta ^2(T_2)/(E{E}') \right ]$, $\Sigma$ is the electron-phonon coupling constant, and $\zeta$ is the Riemann zeta function. We are assuming that the lattice phonons are very well thermalized with the substrate that resides at $T_{bath}$, thanks to the vanishing Kapitza resistance between thin metallic films and the substrate at low temperatures~\cite{Wel94,Gia06}. 

Going forward in the description of the terms in Eq.~\eqref{ThermalBalanceEq}, $c_v(T)=T\frac{\mathrm{d} \mathcal{S}(T)}{\mathrm{d} T}$ is the volume-specific heat capacity, with $\mathcal{S}(T)$ being the electronic entropy density of the superconductor $S_2$~\cite{Rab08,Sol16}
\begin{eqnarray}
&&\mathcal{S}(T)=-4k_BN_F\int_{0}^{\infty}d\varepsilon \mathcal{N}_2(\varepsilon,T)\times \\\nonumber
&&\times\left\{ \left [ 1-f(\varepsilon,T) \right ] \log\left [ 1-f(\varepsilon,T) \right ]+f(\varepsilon,T) \log f(\varepsilon,T)\right \}.\nonumber
\label{Entropy}
\end{eqnarray}
In Eq.~\eqref{ThermalBalanceEq}, $\kappa(T_2)$ is the electronic heat conductivity, given by~\cite{For17}
\begin{equation}\label{electronicheatconductivity}
\kappa(T_2)=\frac{\sigma_N}{2e^2k_BT_2^2}\int_{-\infty}^{\infty}\mathrm{d}\varepsilon\varepsilon^2\frac{\cos^2\left \{ \text{Im} \left [\text{arctanh} \left (\frac{\Delta(T_2)}{\varepsilon+i\gamma_2} \right )\right ] \right \}}{\cosh ^2 \left (\frac{\varepsilon}{2k_BT_2} \right )}.
\end{equation}
\begin{figure}[!!t]
\includegraphics[width=\columnwidth]{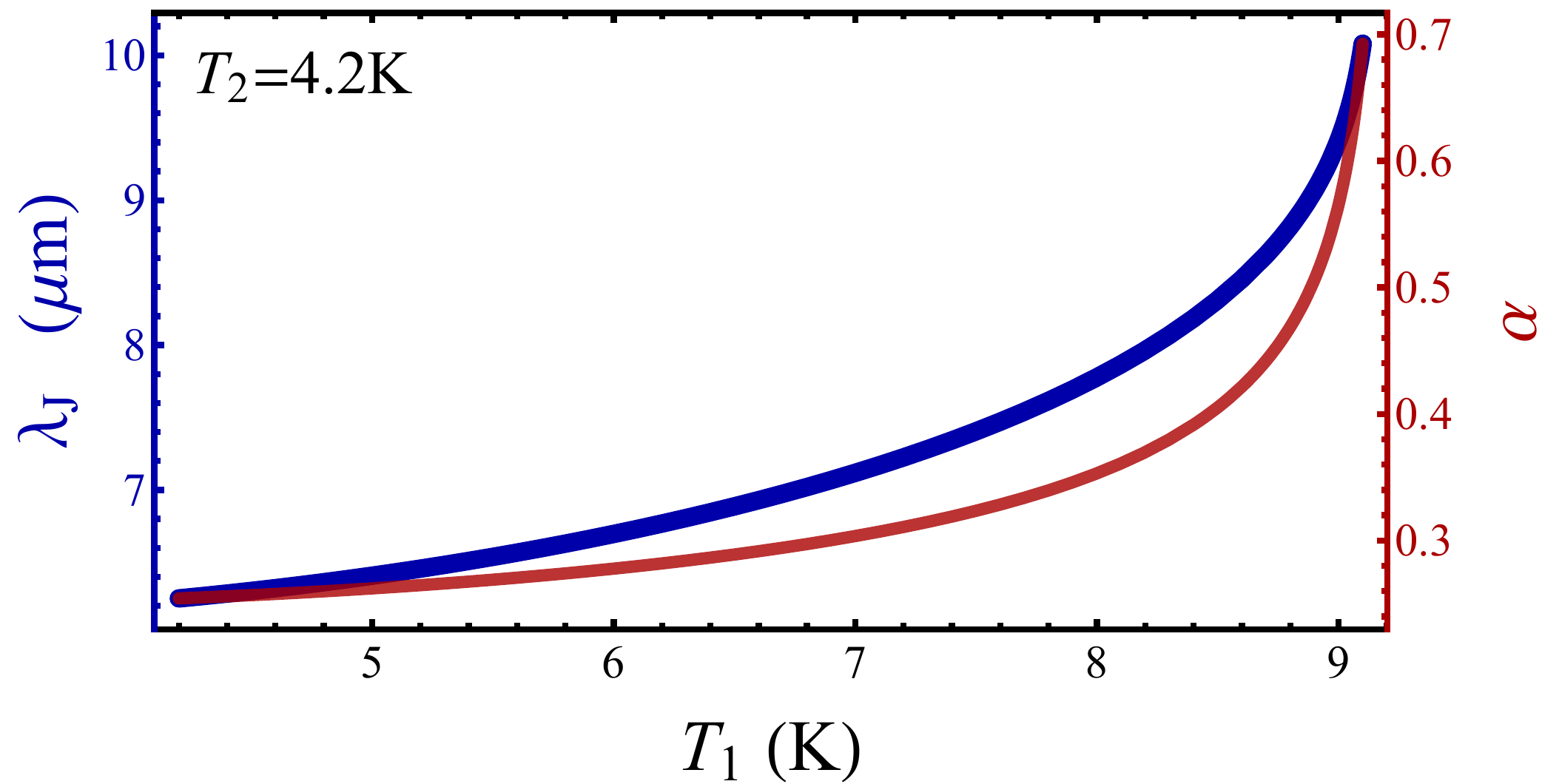}
\caption{Josephson penetration length $\lambda_{_{J}}$ (left vertical scale, blu line) and damping parameter $\alpha$ (right vertical scale, red line) as a function of the temperature of the hot electrode $T_1$, for $T_2=4.2\;\text{K}$, for a Nb-based LJJ with values of the junction parameters discussed in the main text.}
\label{Fig03}
\end{figure}

In order to comprehensively account all the thermal effects, we observe also that the temperature affects both the effective magnetic thickness $t_d(T_1,T_2)$ and the Josephson critical current $I_c( T_1,T_2)$, which varies with the temperatures according to the generalized Ambegaokar and Baratoff formula~\cite{Gia05,Tir08,Bos16}
\begin{eqnarray}\label{IcT1T2}\nonumber
I_c ( T_1,T_2 )=&&\frac{1}{2eR}\Bigg|\int_{-\infty}^{\infty} \Big\{ f( \varepsilon ,T_1 )\textup{Re}\left [\mathfrak{F}_1(\varepsilon ) \right ]\textup{Im}\left [\mathfrak{F}_2(\varepsilon ) \right ]\\
&&+ f( \varepsilon ,T_2 )\textup{Re}\left [\mathfrak{F}_2(\varepsilon ) \right ]\textup{Im}\left [\mathfrak{F}_1(\varepsilon ) \right ] \Big\} d\varepsilon \Bigg|,
\end{eqnarray}
where $\mathfrak{F}_j(\varepsilon ) =\Delta_j \left ( T_j \right )\Big/\sqrt{\left ( \varepsilon +i\gamma_j \right )^2-\Delta_j^2 \left ( T_j\right )}$.
Accordingly, both the Josephson penetration depth $\lambda_{_{J}}$ and the damping parameter $\alpha$ vary with the temperatures, see Fig.~\ref{Fig03}. Since the soliton width depends on $\lambda_{_{J}}$, this thermal dependence affects both the dynamics and the shape of the soliton and, then, the temperature profile along the junction. 

The feasibility to affect the soliton dynamics by locally heating the system is the cornerstone of the low temperature scanning electron microscopy (LTSEM)~\cite{Lac93,Gro94,Mal94,Dod97}. This techniques was proved to be a powerful experimental tool for investigating fluxon dynamics in Josephson devices. The main idea behind this technique is to locally heat a small area ($\sim\mu\text{m}$) of the junction by a narrow electron beam. The generated hot spot acts as a small thermal perturbation with the aim to drastically locally increase the effective dissipation coefficient. This process results in a change of the I-V characteristic of the device. By gradually scanning the electron beam along the junction surface and measuring the voltage, an ``image'' of the dynamical state of the LJJ can be produced. Alternatively, in our work we discuss a sort of thermal imaging of a magnetically excited soliton, through the temperature profile of the floating electrode of the device. 

In Fig.~\ref{Fig03}, we assume a fixed $T_2$, since, in the small range of variation of $T_2$ that we will discuss, the effect of this temperature on $\lambda_{_{J}}$ and $\alpha$ is vanishingly small, and then can be neglected. 

Finally, we assume that the electrode $S_2$ is initially at $T_2(x,0)=T_{bath}\;\forall x\in[0,L]$, and that its ends are thermally isolated, so that boundary conditions of Eq.~\eqref{ThermalBalanceEq} read $\left . \frac{\partial T_2}{\partial x} \right |_{x=0,L}=0$. 
The choice of the initial temperature of the electrode $S_2$ is not essential for our discussion, since we will assume to excite a soliton only when $T_2$ reaches a steady value $T_{2,s}$ in-between $T_{bath}$ and $T_1$.

\section{Results}
\label{Results}\vskip-0.2cm

We consider an Nb/AlO$_x$/Nb SIS LJJ characterized by a resistance per area $R_a=50~\Omega~\mm^2$ and a specific capacitance $C_s=50~fF/\mu \text{m}^2$. The linear dimensions of the device are $L=150\;\mm$, $W=0.5\;\mm$, $D_2=0.1\;\mm$, and $d=1\text{nm}$. For the Nb electrode, we assume $\lambda_{L}^0=80\text{nm}$, $\sigma_N=6.7\times10^6 \Omega^{-1}\text{m}^{-1}$, $\Sigma=3\times10^9\textup{W}\textup{m}^{-3}\textup{ K}^{-5}$, $N_F=10^{47}\textup{ J}^{-1}\textup{ m}^{-3}$, $\Delta_1(0)=\Delta_2(0)=\Delta=1.764k_BT_c$, with $T_c=9.2\;\text{K}$ being the common critical temperature of the superconductors, and $\gamma_1=\gamma_2=10^{-4}\Delta$.

\begin{figure}[!!t]
\includegraphics[width=\columnwidth]{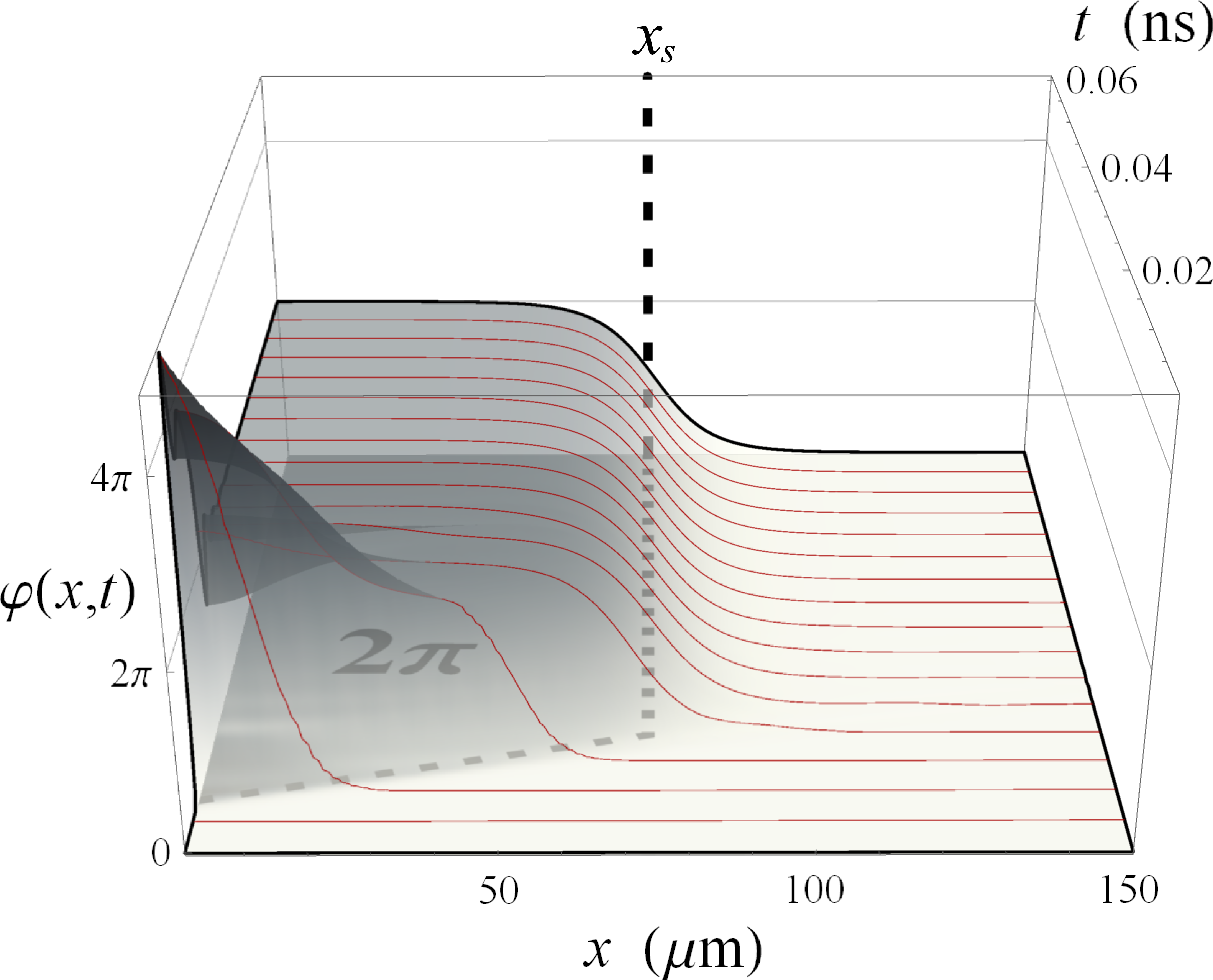}
\caption{Phase evolution as a function of the position $x$ and the time $t$, for $T_1=7\;\text{K}$ and $T_{bath}=4.2\;\text{K}$. A soliton magnetically excited in $x=0$ shifts along the junction. Correspondingly, the Josephson phase $\varphi$ undergoes a 2$\pi$ step (see red lines). The phase values and the position $x_s$ of the soliton, which is marked by a black dashed line, are highlighted in the contour plot underneath the main graph.}
\label{Fig04}
\end{figure}
\begin{figure*}[!!t]
\includegraphics[width=0.75\columnwidth]{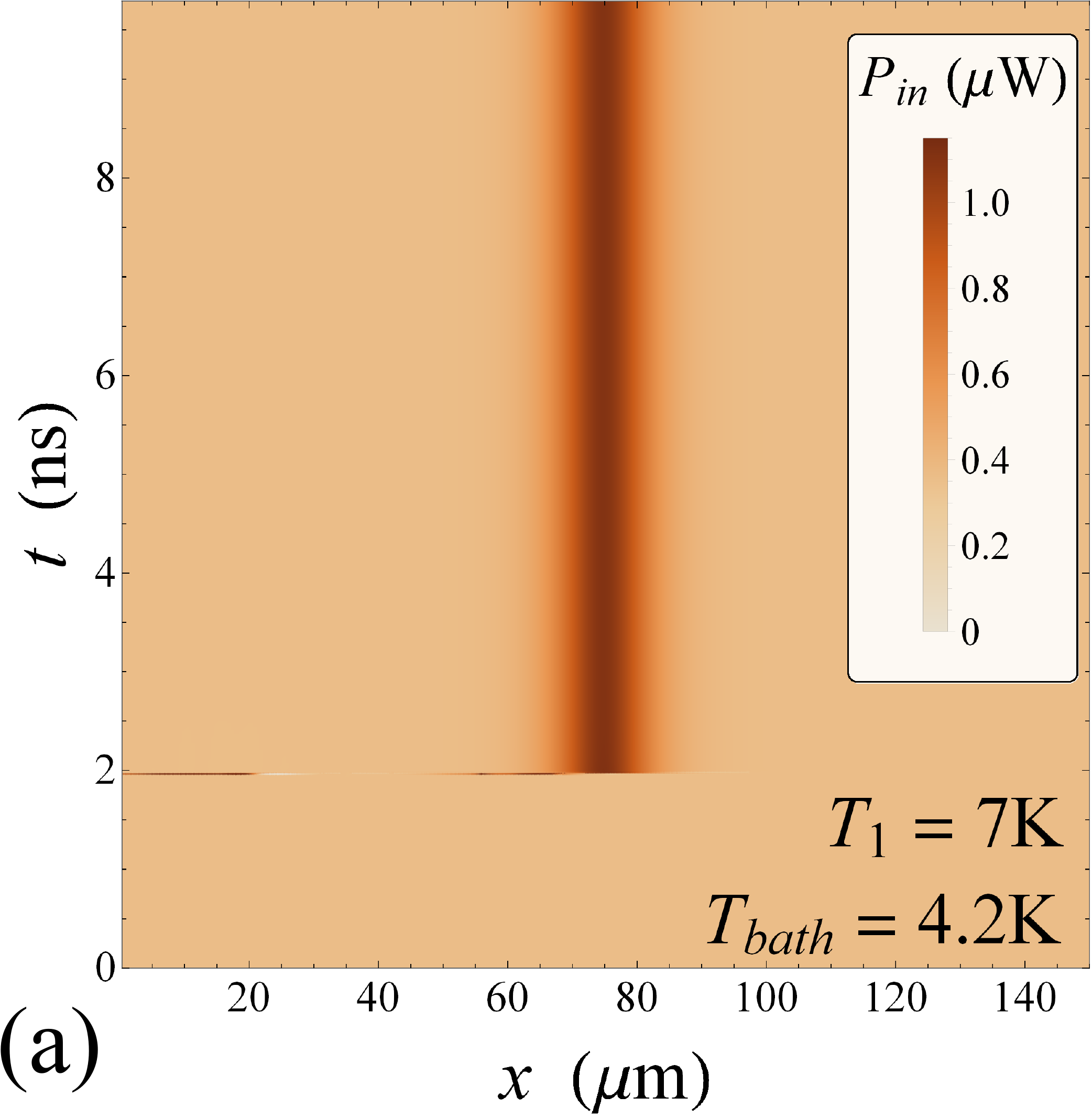}\hspace{0.4cm}
\includegraphics[width=1.25\columnwidth]{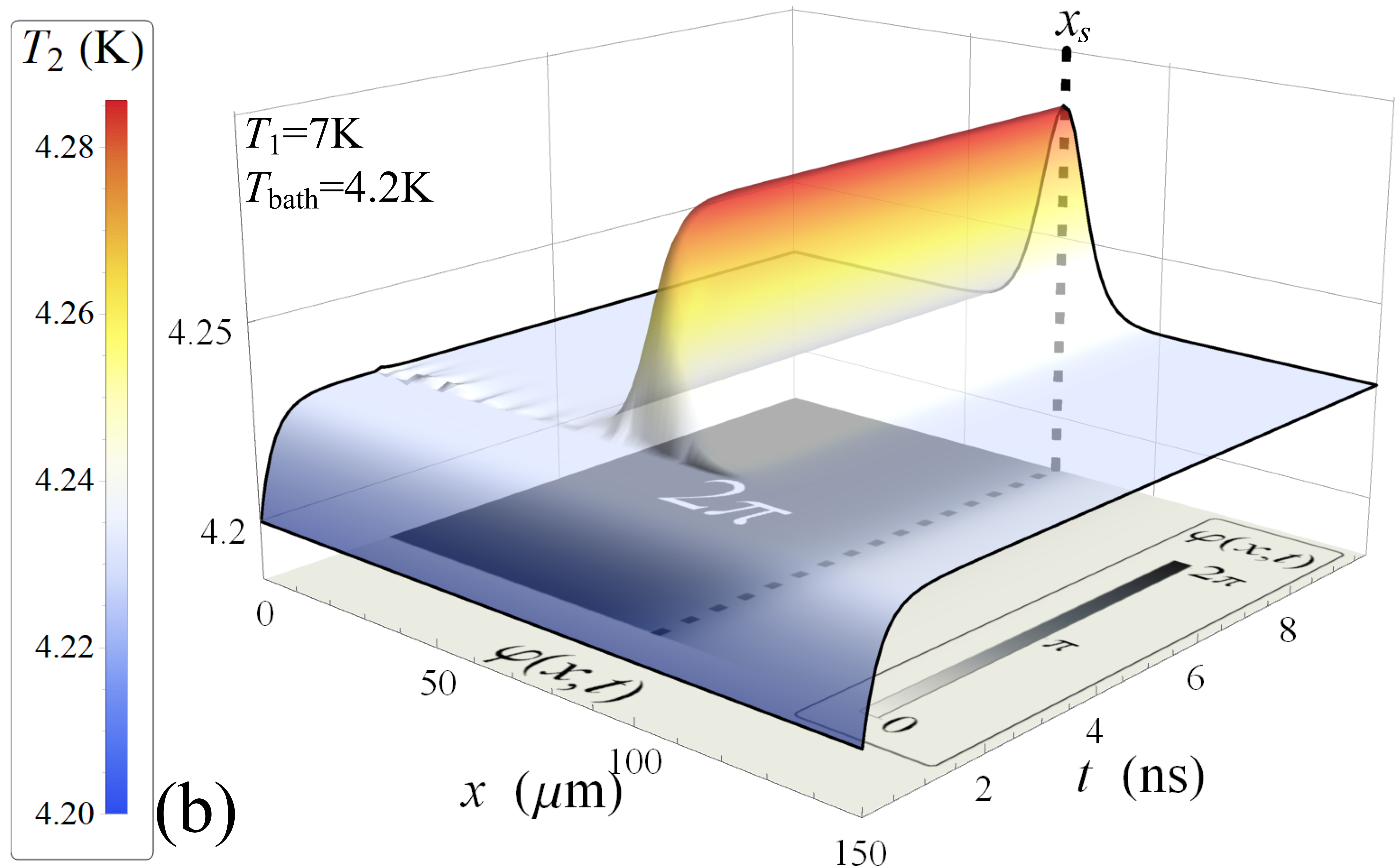}
\caption{\textbf{a}, Heat current $P_{in}( T_1,T_2,\varphi,V)$ flowing from $S_1$ to $S_2$, see Eq.~\eqref{Pt}. \textbf{b}, Evolution of the temperature $T_2(x,t)$ of $S_2$. In both panels, the soliton is magnetically excited to the left end, i.e., $x=0$, after $\sim2\;\text{ns}$. At this time, the superconducting electrode $S_2$ is already fully thermalized at the steady temperature $T_{2,s}\sim4.23\;\text{K}$. Then, in correspondence of the induced soliton, we observe a clear enhancement of both $P_{in}$ and $T_2$. In panel (b), the phase values $\varphi(x,t)$ and the position of the soliton, which is marked by a black dashed line, are highlighted in the contour plot underneath the main graph. For both panels, $T_1=7\;\text{K}$, $T_{bath}=4.2\;\text{K}$, and the junction is initially at the temperature $T_2(x,0)=T_{bath}\; \forall x\in[0-L]$.}
\label{Fig05}
\end{figure*}

Here we focus on the simplest case in which we magnetically excite a soliton which then moves along the junction as the friction affecting its dynamics stops it. The resulting standing soliton is stable and, if it is far enough to the junction edges and in absence of further perturbations, definitively remains in this position. Then, to model this situation, the ``left'', i.e., in $x=0$, junction edge is excited by a Gaussian magnetic pulse, with normalized amplitude $ \mathcal{H}_{max}=8.5$ and width $\sigma=1$ (in units of $\frac{\mu_0}{2\pi}\frac{\Phi_0}{t_d\lambda_{_{J}}}$ and $\omega_p^{-1}$, respectively), which induces a soliton moving rightward along the junction. 
The width and the velocity of the generated soliton directly depend on the temperatures of the system through $\lambda_{_{J}}$ and $\alpha$, respectively. In fact, the higher the temperatures the larger both $\lambda_{_{J}}$ and $\alpha$, since both are proportional to $I_c^{-1/2}$ (see Fig.~\ref{Fig03}). Therefore, by increasing the temperatures, the soliton enlarges and slows down, since both $\lambda_{_{J}}$ and $\alpha$ increase. This shows that the manipulation of the thermal profile along the junction can be also eventually used to modify the soliton dynamics~\cite{Kra97}.

We impose a thermal gradient across the system, specifically, the bath resides at $T_{bath}=4.2\;\text{K}$, and $S_1$ is at a temperature $T_1=7\;\text{K}$ kept fixed throughout the computation. The electronic temperature $T_2(x,t)$ of the electrode $S_2$ is the key quantity to master the thermal route across the junction, since it floats and can be driven by controlling the soliton along the system. 

The evolution of the Josephson phase $\varphi(x,t)$ in the presence of a magnetically excited soliton is shown in Fig.~\ref{Fig04}. In this figure, a rightwards moving soliton (which corresponds to a $2\pi$ step of the phase along the junction) at different instants is outlined by red lines, whereas a dashed line in the contour plot underneath the main graph marks the soliton position. As expected, due to the friction (which is accounted by a value of the damping parameter $\alpha=(\omega_p RC)^{-1}\simeq0.3$) the soliton sets in $x_s\sim74.8\;\mm$ and definitively stays in this position.

We observe that in correspondence of the soliton, the heat flux $P_{in}$ clearly enhances (see Fig.~\ref{Fig05}a). Specifically, the steady value of the heat current in correspondence of the soliton is $P_{in}\sim1.1\;\mu\text{W}$, whereas it is $P_{in}\sim0.3\;\mu\text{W}$ elsewhere. 

Finally, the behaviour of the temperature $T_2(x,t)$ reflects the behavior of the thermal flux $P_{in}$, as it is shown in Fig.~\ref{Fig05}b. In this case, the soliton is excited after $\sim2\text{ns}$, namely, as the whole electrode $S_2$ is thermalized at the steady ``unperturbed'' (i.e., unaffected by excitations) temperature $T_{2,s}\simeq4.23\;\text{K}$. Interestingly, the soliton induces a local intense warming-up in $S_2$, with a steady maximum temperature $T_{2,Max}\simeq4.29\;\text{K}$. 

We observe that, as the soliton sets in $x_s$, the temperature enhances exponentially approaching its steady value, see Fig.~\ref{Fig05}b. The thermal response time can be estimate as the characteristic time of the exponential evolution by which the temperature approaches its stationary value. Then, from Fig.~\ref{Fig05}b we deduce the value $\tau_{th}\sim0.25\;\text{ns}$. Markedly, a quite good estimate of this thermal response time results also in a linear response regime, namely, by first order expanding the heat current terms in Eq.~\eqref{ThermalBalanceEq}. In fact, by following the same procedure developed in Ref.~\cite{,GuaSolBra18}, we obtain a thermal switching time $\tau_{sw}\simeq0.1\;\text{ns}$.

\begin{figure}[!!t]
\includegraphics[width=\columnwidth]{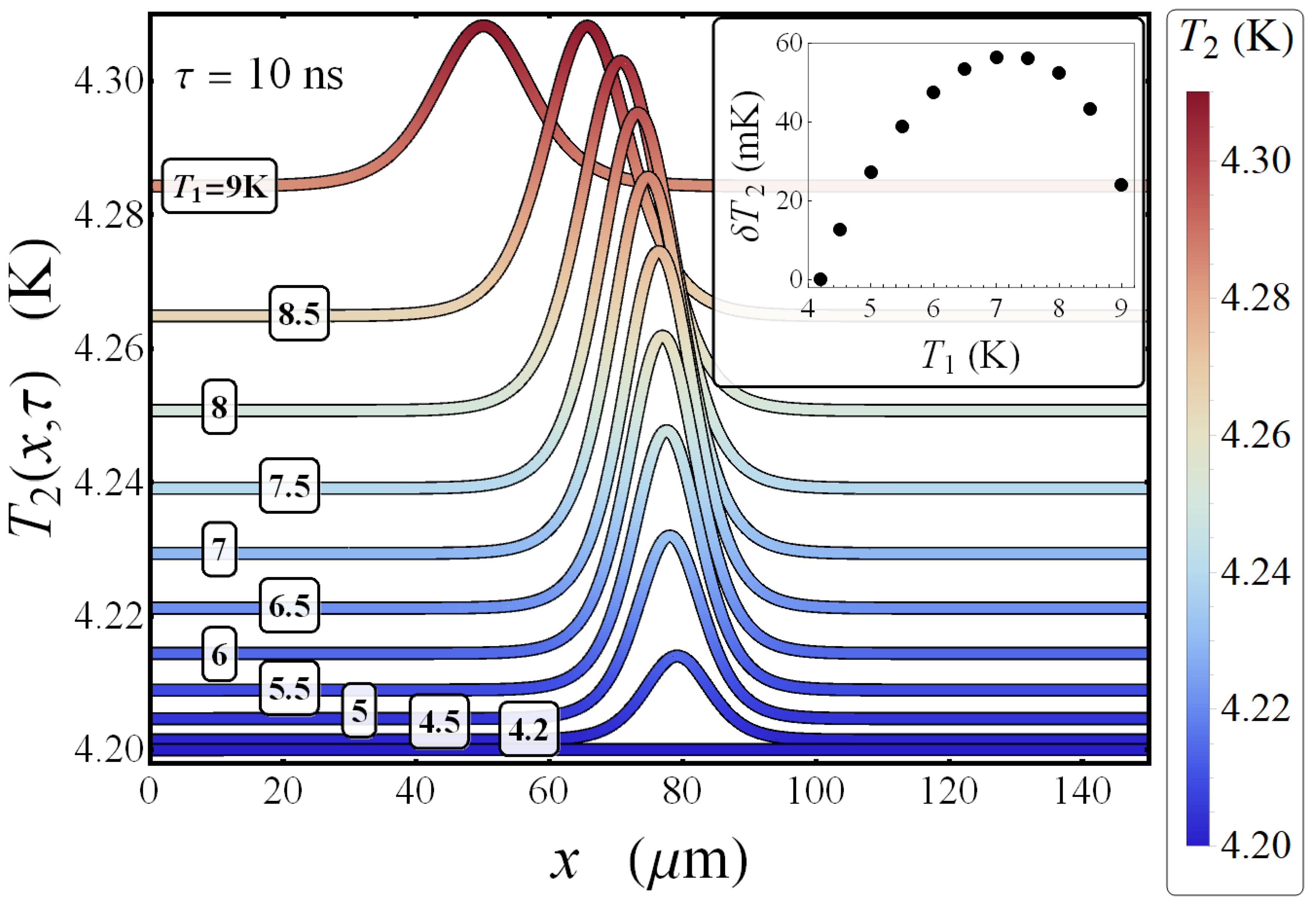}
\caption{Temperature $T_2(x,\tau)$ at $\tau=10\;\text{ns}$ for a few values of $T_1$. In the inset, the $T_2$ modulation amplitude, $\delta T_2$, as a function of $T_1$ is shown. The bath temperature is $T_{bath}=4.2\;\text{K}$ and the junction is initially at the temperature $T_2(x,0)=T_{bath}\; \forall x\in[0-L]$.}
\label{Fig06}
\end{figure}

The role of the temperature $T_1$ is illustrated in Fig.~\ref{Fig06}, where $T_2(x,\tau)$ is calculated at $\tau=10\;\text{ns}$ at a few values of $T_1$ and $T_{bath}=4.2\;\text{K}$. By rising $T_1$, the temperature peak shifts leftwards and becomes wider, just because the soliton slows down and enlarges, as a consequence of the parameter variations discussed in Fig.~\ref{Fig03}. Interestingly, the $T_2$ modulation amplitude, $\delta T_2=T_{2,Max}-T_{2,s}$, defined as the difference between the maximum and the minimum values of $T_2(x,\tau)$ along the junction at a fixed time $\tau$, behaves nonmonotonically by varying $T_1$ (see the inset of Fig.~\ref{Fig06}). In fact, $\delta T_2$ is vanishing for low $T_1$'s (specifically, for $T_1=T_{bath}$ there is no thermal gradient across the system). It then increases up to $\delta T_2\sim56\;\text{mK}$ for $T_1=7\;\text{K}$, and it finally reduces again for $T_1\to T_c$, due to the temperature-induced suppression of the energy gaps in the superconductors.

The physical effect we have described here can promptly find an application as a Josephson thermal router~\cite{Tim17}. Specifically, we can design to direct through a soliton the heat to a superconducting finger electrode, attached for instance in $x_s$, in order to selectively warm it up. 
Additionally, this idea can be improved further by including an external electric bias current across the junction. In fact, a bias current density, $J_b$, acts on the soliton with a Lorentz force, $\mathbf{F}_L=\mathbf{J}_b\times\mathbf{\Phi}_0$ (with the direction of $\mathbf{\Phi}_0$ depending on the polarity of the soliton).
So, in the presence of an external bias current, according to the perturbational approach~\cite{McL78}, a soliton drifts with a velocity approximately given by $u_{d}=1\Big /\sqrt{1+\left [ 4\alpha/\left ( \pi \gamma  \right ) \right ]^2}$~\cite{Ust98}, with $\gamma=J_b/J_c$. Specifically, for a low bias current $u_d\simeq \frac{\pi\gamma}{4\alpha}$. This allows us to actively control the dynamics and the final position of the soliton and, thus, the local temperature of the electrode.
Therefore, a multi-terminal device allowing to distribute the heat among several reservoirs can be conceived, in which we can select which terminal to heat by shifting through the bias current the soliton along the junction. Clearly, the time dependent approach we illustrated so far is indispensable to accurately describe the dynamical temperature response when the soliton moves from a finger to the next one, and then to properly master the operating principles of a multi-terminal device.

\section{Conclusions}
\label{Conclusions}\vskip-0.2cm
In conclusion, we have discussed the phase-coherent thermal transport in a temperature-biased LJJ, where the thermal conduction across the system can be controlled through solitonic excitations. Specifically, we analyse the evolution of the temperature $T_2$ of the floating ``cold'' electrode of the junction, as the temperature $T_1$ of the ``hot'' electrode is kept fixed and the thermal contact with a phonon bath is taken into account. Specifically, in correspondence of a magnetically excited soliton we observe a clear enhancement of the heat current $P_{in}$ flowing through the junction. Correspondingly, a soliton-induced temperature peak occurs, with height up to $\delta T_2\sim56\;\text{mK}$ in a realistic Nb-based proposed setup.

Finally, the physical properties of the device depend on the evolution of the superconducting order parameter along the junction, and, hence, on the dynamics of solitons which can be accurately controlled by external magnetic field, bias current, and shape engineering. This flexibility will allow to suggest new caloritronics applications enabling, for instance, the handling of the local thermal transport in specific points of the junction, i.e., a solitonic thermal router. The analysis shows also the possibility to affect the solitonic properties by manipulating the thermal profile, increasing the possible interplay between thermal and solitonic dynamics. Additionally, the solitonic nature of the system ensures the protection against environmental disturbances and a highly-controllable, unaffected by noise, heat flow. The results obtained will clarify the interplay between solitons and caloritronics at nanoscale, paving the way to the realization of new coherent devices based on the soliton-sustained thermal transport.

Moreover, this device could represent the link between two recent proposals concerning a Josephson based phase-tunable thermal logic~\cite{Pao18} and a logic using fluxons in LJJs~\cite{Wus18}.

The suggested systems could be implemented by standard nanofabrication techniques through the setup used, for instance, for the short JJs-based thermal diffractor~\cite{Mar14}. The modulations of the temperature of the drain ``cold'' electrode is usually obtained by realizing a Josephson junction with a large superconducting electrode, which temperature is blocked at a fixed value, and a small electrode with a small thermal capacity. In this way, the heat transferred significantly affects the temperature of the latter electrode, which is then measured. 

\begin{acknowledgments}
C.G. and F.G. acknowledges the European Research Council under the European Union's Seventh Framework Program (FP7/2007-2013)/ERC Grant agreement No.~615187-COMANCHE for partial financial support. 
P.S. has received funding from the European Union FP7/2007-2013 under REA Grant agreement No. 630925 -- COHEAT and from MIUR-FIRB2013 -- Project Coca (Grant No.~RBFR1379UX). 
A.B. acknowledges the Italian’s MIUR-FIRB 2012 via the HybridNanoDev project under Grant no. RBFR1236VV and CNR-CONICET cooperation programme ``Energy conversion in quantum nanoscale hybrid devices''.
\end{acknowledgments}



%

\end{document}